\documentclass[aps,twocolumn]{revtex4}

\usepackage{graphicx}% Include figure files
\usepackage{dcolumn}% Align table columns on decimal point
\usepackage{bm}% bold math
\usepackage{amsmath}
\begin{document}

\newcommand{\bk}{{\bf k}}
\newcommand{\bp}{{\bf p}}
\newcommand{\bv}{{\bf v}}
\newcommand{\bq}{{\bf q}}
\newcommand{\bQ}{{\bf Q}}
\newcommand{\br}{{\bf r}}
\newcommand{\bR}{{\bf R}}
\newcommand{\bB}{{\bf B}}
\newcommand{\bA}{{\bf A}}
\newcommand{\bK}{{\bf K}}
\newcommand{\vd}{{v_\Delta}}
\newcommand{\tr}{{\rm Tr}}
\newcommand{\kslash}{\not\!k}
\newcommand{\qslash}{\not\!q}
\newcommand{\pslash}{\not\!p}
\newcommand{\rslash}{\not\!r}
\newcommand{\bs}{{\bar\sigma}}

\title{Gauge invariant response functions in Algebraic Fermi liquids}

\author{M. Franz\rlap,$^1$ T. Pereg-Barnea\rlap,$^1$ D. E. Sheehy$^1$ and 
Z. Te\v{s}anovi\'c$^2$}
\affiliation{$^1$Department of Physics and Astronomy,
University of British Columbia, Vancouver, BC, Canada V6T 1Z1\\
$^2$Department of Physics and Astronomy,
Johns Hopkins University, Baltimore, MD 21218}

\date{\today}

\begin{abstract}
A new method is developed that permits the simple evaluation of two-loop
response functions for fermions coupled to a gauge field. 
We employ this method to study the gauge-invariant 
response functions in the Algebraic Fermi liquid, a non-Fermi
liquid state proposed to describe the pseudogap phase
in the QED$_3$ theory of cuprate superconductors.
The staggered spin susceptibility is found to exhibit a characteristic 
anomalous dimension exponent $\eta_4$, while other correlators show
behavior consistent with the conservation laws imposed by 
the symmetries of the underlying theory.

\end{abstract}

\maketitle
%\pacs{74.60.-w,74.60.Ec,74.72.-h}

Low-energy effective theories of certain correlated electronic systems are 
known to be equivalent to
(2+1) dimensional quantum electrodynamics (QED$_3$) 
\cite{semenoff1,haldane1,mavromatos1,lee1,rantner1,ft1}. 
The latter describes $N$ species 
of massless `relativistic' Dirac fermions coupled to a massless
gauge field $a_\mu$. While the physics behind these different reincarnations
of QED$_3$ varies from case to case, 
these theories are all of considerable general interest for the following 
reason: In
the so-called symmetric phase of QED$_3$, long-range interactions 
mediated by the massless gauge field produce characteristic 
anomalous correlations between electrons which decay on long length- and 
time-scales as {\em nontrivial power laws}. This symmetric phase has been 
termed variously as the Algebraic Spin~\cite{rantner1} or Algebraic 
Fermi~\cite{ft1} liquid (AFL) and embodies a unique realization of a 
non-Fermi liquid state of electronic
matter in 2 spatial dimensions.

The power law correlations are encoded in the fermion propagator of the
theory, $G(x)= \langle\Psi(x)\bar{\Psi}(0)\rangle 
\sim x^{-(2+\eta)}$, where $\eta=(4/3\pi^2 N)(3\xi-2)$ is
the anomalous dimension exponent and $\xi\geq 0$ is 
a gauge fixing parameter.
The above fermionic propagator is gauge-dependent and therefore
it cannot represent the behavior of the physical electron in the
underlying theory. 
Much effort has gone into 
constructing and calculating the proper gauge-invariant electron 
propagator~\cite{khvesh2,rantner2,ftv1,gusynin1} but the situation remains unclear.
The most natural candidate, ${\cal G}(x-x')=\langle\exp(i\int_x^{x'}a_\mu
ds_\mu)\Psi(x)\bar{\Psi}(x')\rangle$, suffers from severe ultraviolet
divergences due to the line integral of the gauge field and is 
meaningless in the absence of a physically motivated regularization 
scheme~\cite{ftv1}. Attempts to evaluate ${\cal G}$ directly have yielded
an unphysical negative anomalous dimension~\cite{khvesh2,rantner2,gusynin1}.

In this Letter we search for anomalous power law correlations
in various physical response functions of QED$_3$
which represent susceptibilities and conductivities of the spin and
charge degrees of freedom in the underlying theory. By construction these 
quantities describe gauge invariant physical responses of the system to 
an external probe and therefore no difficulties arise
as to the interpretation of the results. 
We evaluate the requisite correlators
to leading nontrivial order in the $1/N$ expansion \cite{REF:AH}. 
The principal technical 
hurdle in any such calculation is the evaluation of the vertex correction
which is essential to preserve gauge invariance \cite{kim1}.
In the typical treatment \cite{chen1,rantner3} this leads to lengthy 
algebra due to overlapping singularities within Feynman diagrams.
Here we devise a new method for isolating 
the leading divergent behavior of such diagrams which reduces 
the task to
computing the trace of small number of Dirac gamma matrices. We test this new
method against known results and derive new results with essentially 
no extra effort. We formulate a simple general rule for determining whether 
or not a given correlator exhibits any anomalous dimension and discuss 
this result in terms of a theorem that prohibits the 
correlators of conserved operators from acquiring  an anomalous dimension~\cite{gross1,wen1}.

In what follows we focus on the QED$_3$ theory of cuprate pseudogap
as formulated in Refs.~\cite{ft1,ftv1} but our technique remains applicable
to any other reincarnation of QED$_3$~\cite{mavromatos1,lee1,rantner1}.
In the former QED$_3$ emerges as a low-energy effective theory for the 
nodal topological fermions $\Psi_l(x)$, in a 
phase-disordered $d$-wave superconductor ($d$SC). The Lagrangian density is 
\begin{equation}
{\cal L}_D=\sum_{l=1}^{2} 
\bar\Psi_l(x) \gamma_\mu (i\partial_\mu -a_\mu)\Psi_l(x) + 
{1\over 2}K_\mu(\partial\times a)_\mu^2,
\label{l1}
\end{equation}
where the gauge field $a_\mu$ encodes the topological frustration 
encountered by fermions as they propagate through the ``soup'' of 
fluctuating unbound vortex--antivortex pairs. $x=(\tau,{\br})$ denotes 
the space-time coordinate and 
$\gamma_\mu$ ($\mu=0,1,2$) are $4\times 4$ gamma matrices satisfying 
$\{\gamma_\mu,\gamma_\nu\}=2\delta_{\mu\nu}$. The bare dynamics of the
gauge field is Maxwellian \cite{ft1,ftv1,herbut1} with stiffness $K_\mu$.

In the above theory (\ref{l1}) $\Psi_l(x)$ is a four component spinor 
describing the fermionic excitations at the $l$-th pair of antipodal 
nodes of the
underlying $d$SC. Its individual components are related to the original 
electron operators $c_\sigma$ through the singular gauge transformation 
\begin{equation}
c_\sigma(\br,\tau)=e^{i\varphi_\sigma(\br,\tau)}
\psi_\sigma(\br,\tau) ,
\label{ft}
\end{equation}
detailed in Refs.~\cite{ft1,ftv1}. The purpose of this transformation is 
 to ``unwind'' the phase $\varphi(\br,\tau)= 
\varphi_\uparrow(\br,\tau)+\varphi_\downarrow(\br,\tau)$  of the fluctuating
SC order parameter $\Delta(\br,\tau)=\Delta_0e^{i\varphi(\br,\tau)}$ 
in favor of coupling to the gauge field $a_\mu$ which 
is related to coarse-grained 
gradients of phases $\varphi_\sigma$. The difficulty in computing the
gauge-invariant electron propagator ${\cal G}(x-x')$
stems from the necessity to evaluate the averages of such
phase factors. 
The latter are dominated by short length-scale physics that is not properly 
described by the effective theory (\ref{l1}). 
\begin{figure}
\includegraphics[clip,width=8.5cm]{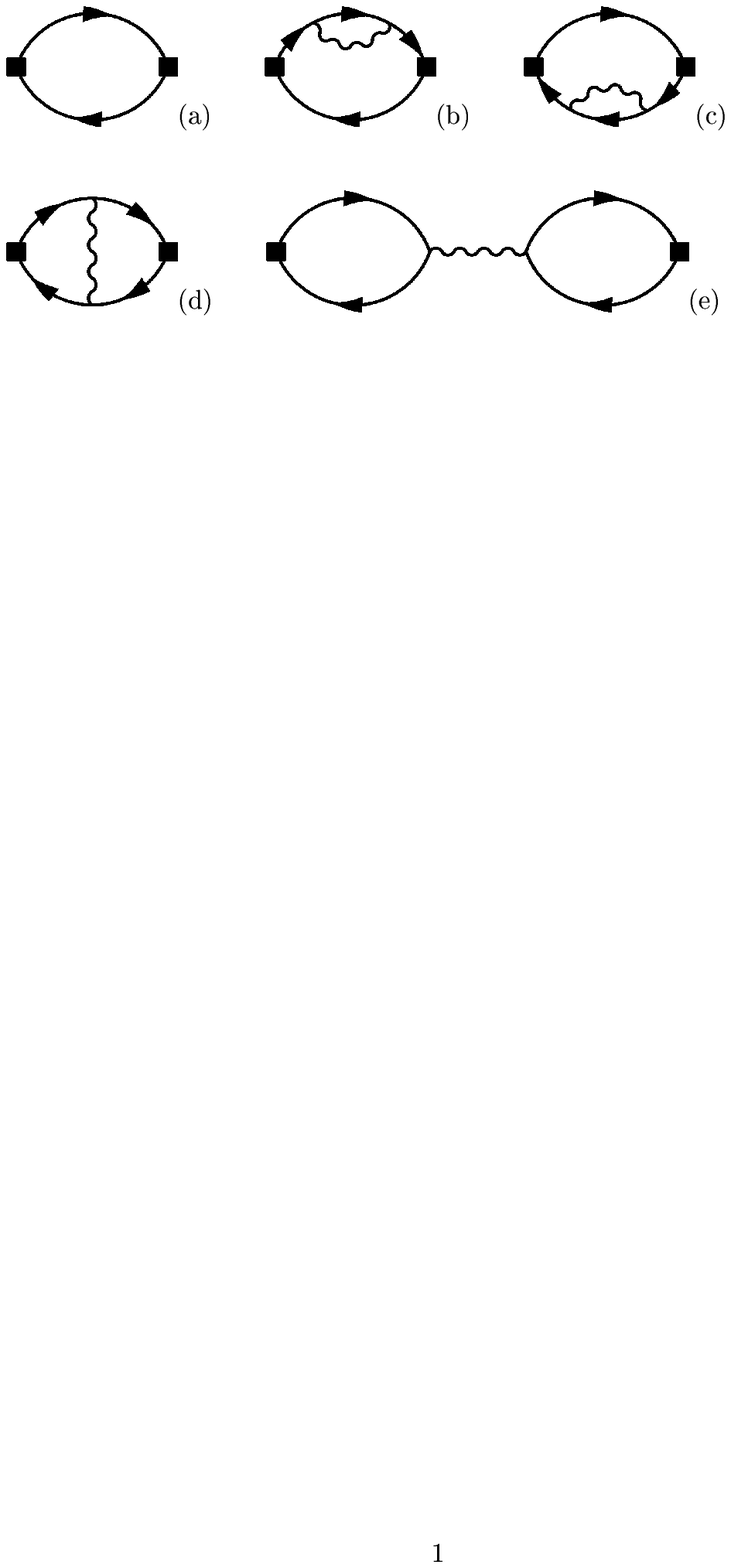}
\caption{\label{fig1}
Feynman diagrams for susceptibility Eq.~(\ref{susc}) to leading 
order in $1/N$. Solid lines represent the bare fermion propagator, wavy
lines represent the screened gauge-field propagator, and the solid squares 
stand for vertex $\Gamma_l$. 
}
\end{figure}

Here we focus on the charge and
spin density correlators which do not suffer from the above problem.
Consider in particular the spin susceptibility,
\begin{equation}
\chi^s(\bq,i\omega)=\int_0^\beta d\tau e^{i\omega\tau}\langle {\rm T}_\tau
S^z_\bq(\tau) S^z_{-\bq}(0)\rangle,
\label{spins}
\end{equation}
with $\beta=1/T$ and $S^z_\bq=\int d^2r e^{i\bq\cdot\br}\sum_\sigma\sigma
c^\dagger_\sigma(\br)c_\sigma(\br)$ the $z$-component of the 
electron spin 
operator at wavevector $\bq$. The charge susceptibility $\chi^c$ is
defined as in Eq.\ (\ref{spins}) but with $S^z_\bq$ replaced by the charge 
density operator $\rho_\bq=\int d^2r e^{i\bq\cdot\br}\sum_\sigma
c^\dagger_\sigma(\br)c_\sigma(\br)$. Both spin and charge densities are 
gauge singlets with respect to $a_\mu$, i.e., they do not 
pick up any phase factors under the transformation (\ref{ft}). Following
Ref.\ \cite{ftv1} we adopt the representation for the gamma matrices given 
by $\gamma_\mu=\sigma_3\otimes(\sigma_2,\sigma_1,-\sigma_3)$ with $\sigma_i$
the Pauli matrices. We further define $\gamma_3=\sigma_1\otimes\openone$
and $\gamma_5=\sigma_2\otimes\openone$ and note that these anticommute
with all $\gamma_\mu$'s. With these definitions we may express the 
density operators  as
\begin{equation}
\begin{array}{c}
S^z(q)\\
\rho(q)
\end{array}
\biggr\}=\int_p\bar\Psi_l(p)\Gamma_l\Psi_l(q-p),
\label{densities}
\end{equation}
where vertex $\Gamma_l$ takes the form $\Gamma_l^s=(i\gamma_0,i\gamma_0)$
and $\Gamma_l^c=(-i\gamma_0\gamma_2,i\gamma_0\gamma_1)$ for spin and charge
densities with momentum transfer close to $\bq=(0,0)$, respectively,
and $\Gamma_l^s=(\gamma_5,0)$, $\Gamma_l^c=(-\gamma_0\gamma_5,0)$ for 
momentum transfer near $\bq={\bf Q}_1 \approx(\pi,\pi)$.
We have switched to Euclidean notation with 3-momenta $q=(q_0,\bq)$, 
$\int_p$ denotes $\int d^3p/(2\pi)^3$ and summation over $l=1,2$ is assumed.
The susceptibilities are then given as
\begin{equation}
\chi(q)=\int_p\int_{p'}\langle \bar\Psi_l(p)\Gamma_l\Psi_l(p+q)
\bar\Psi_l(p')\Gamma_l\Psi_l(p'-q)\rangle.
\label{susc}
\end{equation}

We evaluate $\chi(q)$ at $T=0$ by formally considering $N/2$ identical copies
of fields $\Psi_1$ and $\Psi_2$. Then 
to  leading order in $1/N$ expansion ~\cite{REF:AH} we require 
4 diagrams depicted in Fig.\ \ref{fig1}(a--d). The wavy line represents the 
gauge field propagator that becomes universal at long wavelengths due
to the screening by topological fermions \cite{ft1},
\begin{equation}
D_{\mu\nu}(q)={8\over qN}\left(\delta_{\mu\nu}-{q_\mu q_\nu\over q^2}
(1-\xi)\right),
\label{prop}
\end{equation}
in the sense that $K_\mu$ drops out and only reappears as an upper cutoff
of the theory, $\Lambda\sim 1/K$.

The bare bubble Fig.\ \ref{fig1}(a) reads
\begin{equation}
\chi_0(q)=\int_p{\rm Tr}[G_0(p)\Gamma_l G_0(p-q)\Gamma_l]
\label{susc0}
\end{equation}
where $G_0(p)=\pslash/p^2\equiv p_\mu\gamma_\mu/p^2$ is the free Dirac
propagator, and can be evaluated by standard methods (see e.g. Appendix B
of Ref.\ \cite{ftv1}). One obtains
\begin{equation}
\chi_0(q)=-{\rm Tr}[\gamma_\mu\Gamma_l\gamma_\nu\Gamma_l]
{q\over 64}\left(\delta_{\mu\nu}+{q_\mu q_\nu\over q^2}\right).
\label{susc00}
\end{equation}
We observe that the structure of the bare bubble susceptibility is 
determined entirely by the commutation relation of the vertex with the 
$\gamma$ matrices. We shall see that this property remains true
for the higher order diagrams as well.

Diagrams (b), (c) and (d) represent the leading $1/N$ corrections due to 
the gauge field and contain all the interesting physics. We note that 
their sum is {\em gauge invariant}. This follows since they represent 
correlators of gauge invariant density operators. One can explicitly 
verify that this is so by employing the Ward identity $G_0^{-1}(p+k)-
G_0^{-1}(k)=p_\mu\gamma_\mu$. We are thus free to fix the gauge and in what 
follows we adopt the Feynman gauge ($\xi=1$) in which $D_{\mu\nu}(q)=(8/N)
\delta_{\mu\nu}/q$.
Diagrams (b) and (c) contain the self-energy insertion and read
\begin{equation}
\chi_{(b+c)}(q)=2\int_p{\rm Tr}[G_0(p)\Gamma_l G_0(p-q)
\Gamma_l G_0(p)\Sigma(p)]
\label{suscbd}
\end{equation}
with $\Sigma(p)=\int_k D_{\mu\nu}(k)\gamma_\mu G_0(p-k)\gamma_\nu$=$
-\eta_F\pslash\ln(\Lambda/p)$, and
\begin{equation}
\eta_F={4\over 3\pi^2N}
\label{etaf}
\end{equation}
the fermion anomalous dimension in the Feynman gauge \cite{ftv1}.
Inserting $\Sigma(p)$ into Eq.\ (\ref{suscbd}) and combining with 
Eq.\ (\ref{susc00}) one can express the divergent contribution as
\begin{equation}
\chi_{(b+c)}(q)\simeq 2\eta_F\chi_0(q)\ln(\Lambda/q).
\label{suscbd1}
\end{equation}

Diagram (d) contains the vertex correction and reads 
\begin{widetext}
\begin{equation}
\chi_{(d)}(q)=\int_k\int_p{\rm Tr}\Bigl([G_0(p-q)\Gamma_l G_0(p)]
\gamma_\mu[G_0(p-k)\Gamma_l G_0(p-k-q)]\gamma_\nu\Bigr)D_{\mu\nu}(k).
\label{suscc1}
\end{equation}
As mentioned above the direct evaluation of this type of integral 
presents a significant challenge \cite{chen1,rantner3}.
We are interested in the leading $q\to 0$ behavior of $\chi(q)$. 
In this limit the $p$ integral has singularities as 
$p\to 0$ and $p\to k$, corresponding to divergences in the first and  
second term in the square brackets respectively. The main contribution to 
the integral therefore comes from the vicinity of these two points and
we may evaluate it by expanding the regular part of the integrand at
the singular point. Retaining only the leading term thus gives
\begin{eqnarray}
\chi_{(d)}(q)&\simeq&\int_k\int_p{\rm Tr}\Bigl(
[G_0(p-q)\Gamma_l G_0(p)]\gamma_\mu[G_0(-k)\Gamma_l G_0(-k-q)]\gamma_\nu
\Bigr)D_{\mu\nu}(k)\nonumber \\
&+&\int_k\int_p{\rm Tr}\Bigl(
[G_0(k-q)\Gamma_l G_0(k)]\gamma_\mu[G_0(p-k)\Gamma_l G_0(p-k-q)]\gamma_\nu
\Bigr)D_{\mu\nu}(k).
\label{suscc2}
\end{eqnarray}
\end{widetext}
Performing a variable shift $p\to p+k$ in the second term 
%and exploiting the cyclical property of the trace 
we may simplify the above expression as
\begin{equation}
\chi_{(d)}(q)=2{\rm Tr}\left(
\Omega_l(q)\int_p[G_0(p-q)\Gamma_l G_0(p)]\right)
\label{suscc3}
\end{equation}
with $\Omega_l(q)=\int_k\gamma_\mu[G_0(k)\Gamma_l G_0(k-q)]\gamma_\nu
D_{\mu\nu}(k)$.
%
%\begin{equation}
%\Omega_l(q)=\int_k\gamma_\mu[G_0(k)\Gamma_l G_0(k-q)]\gamma_\nu
%D_{\mu\nu}(k).
%\label{omeg1}
%\end{equation}
%
The last integral is again easy to evaluate and gives
\begin{equation}
\Omega_l(q)=[\gamma_\mu\gamma_\nu\Gamma_l\gamma_\nu\gamma_\mu]\eta_F
\ln(\Lambda/q).
\label{omeg2}
\end{equation}

We now notice that for vertices composed of products of gamma matrices
(such as those entering the spin and charge densities defined above) 
it holds that
\begin{equation}
\gamma_\mu\gamma_\nu\Gamma_l\gamma_\nu\gamma_\mu=\lambda\Gamma_l,
\label{gamma}
\end{equation}
where $\lambda$ is a number. In particular we shall encounter two types
of vertices. Type-I vertices commute or anticommute with all $\gamma_\mu$'s
(e.g. $\Gamma_l=\openone,\gamma_3,\gamma_5,\gamma_3\gamma_5$) and 
in this case $\lambda=9$. Type-II vertices
anticommute with one or two of $\gamma_\mu$'s and commute with the 
rest (e.q. $\Gamma_l=\gamma_0,\gamma_0\gamma_1,\dots$). In this case
$\lambda=1$. With this insight Eq.\ (\ref{suscc3}) becomes
\[
\chi_{(d)}(q)=2\eta_F\lambda\ln(\Lambda/q)\int_p{\rm Tr}
[G_0(p)\Gamma_l G_0(p-q)\Gamma_l].
\]
%
%where the last integral is equal to $\chi_0(q)$. 
Combining this 
result with Eqs.\ (\ref{suscbd1}) and (\ref{susc0}) we can write the 
result for the full susceptibility to $1/N$ order,
\begin{equation}
\chi(q)=\chi_0(q)[1-2\eta_F(1-\lambda)\ln(\Lambda/q)].
\label{susc22}
\end{equation}
This correction may be interpreted as the leading
term of a power law~\cite{REF:AH}, so that we have
\begin{equation}
\chi(q)\sim\chi_0(q)\left({\Lambda\over q}\right)^{\eta_4}, \ \ 
\eta_4={8(\lambda-1)\over 3\pi^2N}.
\label{susc2}
\end{equation}
The anomalous dimension exponent $\eta_4$ is entirely determined by the 
algebraic properties of the vertex $\Gamma_l$ through Eq.\ (\ref{gamma}). 
In particular for type-I vertex ($\lambda=9$)
\begin{equation}
\eta_4={64\over 3\pi^2 N},
\label{nu}
\end{equation}
%
%i.e. $\eta_4\simeq 1.08, 0.54, 0.36$ for $N=2,4,6$, 
while for type-II vertex $\eta_4=0$. 

Returning back to physics we see that the vertex for staggered spin 
susceptibility is type-I and will therefore exhibit nontrivial anomalous 
dimension exponent Eq.\ (\ref{nu}), in agreement with the result of Ref.\
\cite{rantner3} obtained by laborious direct evaluation of the vertex 
correction. Uniform spin susceptibility has type-II vertex and therefore 
does not acquire anomalous dimension from diagrams (b-d), again in 
agreement with \cite{rantner3}.
In addition both charge susceptibilities are type-II and will
not exhibit anomalous dimension.

Finally we note that if $\Gamma_l$ coincides with one 
of the $\gamma_\mu$'s 
then diagrams of the type shown in Fig.\ \ref{fig1}(e) 
are nonvanishing and must be included in the leading order.
One can show that resummation of diagrams (e) modifies the $\chi_0(q)\sim q$
behavior in Eq.\ (\ref{susc00}) to $q^2$. This can be viewed as another type
of anomalous dimension  due to the coupling to the gauge field and we shall 
discuss it more fully elsewhere.

The formalism we have developed allows us to study spin and charge 
conductivities in the AFL. These can be calculated 
through the Kubo formula 
as $\sigma_{ij}(\omega)=-{\rm Im}\Pi^{\rm ret}_{ij}
(\omega)/\omega$, where $\Pi^{\rm ret}_{ij}(\omega)=
\Pi_{ij}(i\omega\to\omega+i\delta)$ with 
\begin{equation}
\Pi_{ij}(i\omega)=-\int_0^\beta d\tau e^{i\omega\tau}\langle {\rm T}_\tau
j_i(\tau) j_j(0)\rangle
\label{kubo1}
\end{equation}
the current-current correlation function. Indices $i,j=1,2$ label the 
spatial components of the 3-current $j_\mu$. The spin current is given 
as $j^s_\mu(x)=\bar\Psi_l(x)\gamma_\mu\Psi_l(x)$ while the electric current
is given as $j^e_\mu(x)=(\bar\Psi_1\gamma_0\gamma_2\Psi_1-
\bar\Psi_2\gamma_0\gamma_1\Psi_2,-\bar\Psi_1\gamma_1\gamma_2\Psi_1,
-\bar\Psi_2\gamma_1\gamma_2\Psi_2)$. To leading $1/N$ order the computation 
of the correlator (\ref{kubo1}) involves the same diagrams depicted
in Fig.\ \ref{fig1} and we can therefore simply adopt the result obtained 
for the susceptibility in Eq.\ (\ref{susc2}). We see that the vertices
involved in $j^s_\mu$ and $j^e_\mu$ are all type-II and therefore neither
spin nor charge conductivity will exhibit anomalous dimension.

We may conclude that of all quantities considered only the staggered spin
susceptibility will bear the unique signature of the AFL in that its
frequency, momentum and temperature dependence will be controlled by
the anomalous dimension exponent $\eta_4$ given in Eq.\ (\ref{nu}). The other
quantities will behave essentially as if the theory was uncorrelated. 
%Two interesting
%questions thus arise: How can one understand this contrasting behavior?
%Is it possible to find other physical observables that will reflect the
%anomalous correlations characteristic of the AFL?

The fact that some quantities remain essentially unaffected by strong
long range correlations can be traced to a field theoretic 
theorem~\cite{gross1,wen1} which states
that the correlator of conserved currents (i.e. currents satisfying 
$\partial_\mu j_\mu=0$) must exhibit a scaling dimension
which agrees with its engineering dimension.
Physically this means that some 
quantities are constrained by their conservation laws to such a degree that
correlations cannot alter their long-distance scaling behavior. As an 
example consider the spin current $j^s_\mu$. It is a conserved current,
$\partial_\mu j^s_\mu=0$ guaranteed by the gauge invariance of the theory
[i.e.\ invariance under local symmetry 
$\Psi_l(x)\to e^{i\theta(x)}\Psi_l(x)$]. As we have verified
by explicit calculation neither spin conductivity nor uniform spin 
susceptibility exhibit anomalous dimension ($\sim 1/N$) beyond
the $\chi_0\sim q^2$ behavior mentioned above, 
in accord with the theorem. Understanding the behavior
of the charge conductivity is less straightforward as we do not expect the
quasiparticle current to be conserved in a (phase-disordered)
superconductor. Indeed there is no symmetry that would guarantee 
$j^e_\mu$ conservation. However, the theory is known to posses a ``chiral''
symmetry ($\Psi_l\to e^{i\gamma_3\gamma_5\phi_l}\Psi_l$) \cite{tvf1} which 
produces
two conserved currents: $j^{(l)}_\mu=\bar\Psi_l\gamma_\mu(i\gamma_3\gamma_5)
\Psi_l$ (no sum over $l$). By virtue of the identity 
$\gamma_0\gamma_3\gamma_5=-\gamma_1\gamma_2$ these are related to
the spatial components of $j^e_\mu$, explaining the lack of
anomalous dimension in the electrical conductivity. 
%Similarly the charge density
%can be related to other components of $j^{(l)}_\mu$.
%It is interesting to note that internodal scattering, produced e.g. by 
%point-like disorder, will break the above symmetry. Thus, with disorder,
%quasiparticle current will no longer be protected and conductivity could
%develop anomalous scaling dimension. We note that the full charge response
%of the system also involves the condensate degrees of freedom which we do
%ynot explicitly consider here.

In the $4\times 4$ representation of Dirac gamma matrices there are
4 type-I vertices listed below Eq.\ (\ref{gamma}). Thus there
should be 3 other physical observables 
which exhibit anomalous scaling dimension. What are these?
In turns out that $\gamma_3$ corresponds to the staggered spin susceptibility
at the wave vector ${\bf Q}_{\bar 1}=-{\bf Q}_1$, while $\openone$ and 
$\gamma_3\gamma_5$ correspond to susceptibilities to formation of 
subdominant (phase-incoherent) SC order in $s$ and $p$ channels,
respectively. These quantities are members of the QED$_3$ chiral 
manifold~\cite{tvf1,herbut1}, 
which is the manifold of broken-symmetry states occuring 
for $N< N_c\approx 32/\pi^2=3.24$ at zero temperature. 
This phenomenon
is known as spontaneous chiral symmetry breaking~\cite{chiral}.
The anomalous dimension exponent $\eta_4$ that we find
in the above susceptibilities appears to anticipate the transition into the
chiral symmetry broken phase.
Indeed if we combine $\chi_0(q)$ from Eq.\ (\ref{susc00})
with Eq.\ (\ref{susc2}) we find $\chi(q)\sim q^{1-\eta_4}$, implying 
divergent $q\to 0$ susceptibility for $N<64/3\pi^2={2\over 3}N_c$.

In the context of the QED$_3$ theory of cuprates \cite{ft1} we expect
the experimental manifestations of the nontrivial anomalous dimension
to appear in the quantities inhabiting
the QED$_3$ chiral manifold, most prominently the staggered spin 
susceptibility
that is directly measurable by neutron scattering. In addition, a 
similar nontrivial
response should obtain in the charge channel near $(0,\pi)$ as a 
consequence of enlarged chiral manifold discussed in \cite{tvf1}. This 
particular response is unique to the present theory and will not be
present in the SU(2) theory of Ref.\ \cite{rantner3}.

We conclude by observing that the physics of the Algebraic Fermi liquid bears
distinct similarity to that of 1D Luttinger liquids. While the single 
particle properties are distinctly non-Fermi liquid like, only certain
measurable physical responses exhibit the unique fingerprint of the 
AFL in the 
form of anomalous scaling dimension. The method developed here allows us
to easily identify these quantities and helps in our search 
for situations where other quantities might develop anomalous behavior.

The authors are indebted to L. Balents, T. Davis, I. Herbut, Y.-B. Kim,
M. Rozali and O. Vafek for valuable discussions.  
This research was supported in part by NSERC (MF,TPB,DES), 
NSF Grant DMR00-94981 (ZT) and by A. P. Sloan Foundation (MF). The authors 
thank the Aspen Center for Physics for hospitality.

\end{document}